\documentclass[8.5pt,twoside,twocolumn]{article}
\usepackage[super,sort&compress,comma]{natbib} 
\usepackage{mhchem}
\usepackage{amsmath}
 \usepackage{times}
\usepackage{sectsty}
\usepackage{balance} 
\usepackage{comment}
\usepackage{graphicx} %
\usepackage{lastpage}
\usepackage[format=plain,justification=raggedright,singlelinecheck=false,font=small,labelfont=bf,labelsep=space]{caption} 
\usepackage{fancyhdr}
\usepackage{amssymb}
\usepackage{color}
\usepackage{mathtools}

\usepackage{todonotes}

\newcommand{\subf}[2]{%
  {\small\begin{tabular}[t]{@{}c@{}}
  #1\\#2
  \end{tabular}}%
}
\begin{document}

\thispagestyle{plain}
\fancypagestyle{plain}{

\renewcommand{\headrulewidth}{1pt}}
\renewcommand{\thefootnote}{\fnsymbol{footnote}}
\renewcommand\footnoterule{\vspace*{1pt}%
\hrule width 3.4in height 0.4pt \vspace*{5pt}} 
\setcounter{secnumdepth}{5}

\makeatletter 
\def\subsubsection{\@startsection{subsubsection}{3}{10pt}{-1.25ex plus -1ex minus -.1ex}{0ex plus 0ex}{\normalsize\bf}} 
\def\paragraph{\@startsection{paragraph}{4}{10pt}{-1.25ex plus -1ex minus -.1ex}{0ex plus 0ex}{\normalsize\textit}} 
\renewcommand\@biblabel[1]{#1}            
\renewcommand\@makefntext[1]%
{\noindent\makebox[0pt][r]{\@thefnmark\,}#1}
\makeatother 
\renewcommand{\figurename}{\small{Fig.}~}
\sectionfont{\large}
\subsectionfont{\normalsize} 

\fancyfoot{}

\fancyfoot[RO]{\footnotesize{\sffamily{1--\pageref{LastPage} ~\textbar  \hspace{2pt}\thepage}}}
\fancyfoot[LE]{\footnotesize{\sffamily{\thepage~\textbar\hspace{3.45cm} 1--\pageref{LastPage}}}}
\fancyhead{}
\renewcommand{\headrulewidth}{1pt} 
\renewcommand{\footrulewidth}{1pt}
\setlength{\arrayrulewidth}{1pt}
\setlength{\columnsep}{6.5mm}
\setlength\bibsep{1pt}

\twocolumn[
  \begin{@twocolumnfalse}
\noindent\LARGE{\textbf{Coarsening and mechanics in the bubble model for wet foams}}
\vspace{0.6cm}

\noindent\large{\textbf{Kseniia Khakalo,$^{\ast}$\textit{$^{a}$} 
Karsten Baumgarten,\textit{$^{b}$}
Brian P. Tighe\textit{$^{b}$} and
Antti Puisto\textit{$^{a}$}}}
\vspace{0.5cm}

\vspace{0.6cm}

\noindent \normalsize{

Aqueous foams are an important model system that displays coarsening 
dynamics.
Coarsening in dispersions and foams is well understood in the dilute and 
dry limits, where the gas fraction
tends to zero and one, respectively. 
However, foams are known to undergo a jamming transition from a 
fluid-like to a solid-like state at an intermediate gas fraction,$\phi_c$. 
Much less is known about coarsening dynamics in wet foams near 
jamming, and the link to mechanical response, if any, remains poorly 
understood.
Here, we probe coarsening and mechanical response using numerical 
simulations of a variant of the Durian bubble model for wet foams 
\cite{gardiner2000coarsening}.
As in other coarsening systems we find a steady state scaling regime with 
an associated particle size distribution. 
We relate the time-rate of evolution of the coarsening process to the 
wetness of the foam and identify a characteristic coarsening time that 
diverges approaching jamming. 
We further probe mechanical response of the system to strain while 
undergoing coarsening. 
There are two competing time scales, namely the coarsening time and the 
mechanical relaxation time. 
We relate these to the evolution of the elastic response and the 
mechanical structure.

}
\vspace{0.5cm}
 \end{@twocolumnfalse}
  ]

\footnotetext{\textit{$^{a}$~Aalto University, School Science, Laboratory of Applied Physics B.O.B 11100, FI-00076 AALTO, Finland; E-mail: Kseniia.Khakalo@aalto.fi}}
\footnotetext{\textit{$^{b}$~Delft University of Technology, Process \& Energy Laboratory, Leeghwaterstraat 39, 2628
CB Delft, The Netherlands}}

\section{Introduction}
Foams are composed of repulsively interacting gas bubbles dispersed in a 
liquid phase. Based on gas fraction they are categorized 
as ``wet'' or ``dry''~\cite{weaire}. In wet foams, the liquid concentration is 
higher and the bubbles mostly retain their spherical shape. In 
dry foams, only thin films of liquid separate the bubbles due to the limited 
amount of liquid available. Therefore, the bubbles in dry foams appear in 
polyhedral shapes. The mechanical properties and rheology of a foam also
depend on its gas fraction. At gas fractions well below the critical value $\phi_c$ 
($\approx 0.84$ in 2D and 0.64 in 3D), their mechanical response is mainly determined 
 by the background fluid~\cite{weaire}. At gas fractions above $\phi_c$, 
they  form amorphous, jammed assemblies. Beyond this limit, their mechanical
response can be expected to depend on the bubble properties, such as their
interaction potentials and size distribution.

Foams are not thermodynamically stable, and the bubble size distribution evolves due to destabilization of the 
foam. The three principal mechanisms governing foam destabilization are 
drainage, coarsening, and coalescence~\cite{weaire}. Drainage removes liquid 
between the bubbles via gravity or evaporation increasing the gas 
concentration. Coarsening occurs when gas diffuses from smaller bubbles 
to larger ones, thanks to the difference in their internal pressure \cite{isert}. 
Coalescence, where bubbles join when thin films rupture between them,  primarily 
takes place in dry foams.

Self-similar scaling arguments provide valuable insight into the prolonged coarsening observed in foams 
 \cite{weaire,durian_JPHYS,lambert,lambert10}.
In the scaling state, the growth of bubbles is expected to reach an asymptotic
limit with the average radius following the scaling law 
\begin{equation}
\frac{\langle  R \rangle }{\langle  R_{\rm in} \rangle } \simeq \left \lbrace
\begin{array}{cc}
1 & t \ll \tau_c \\
\left( {t}/{\tau_c} \right)^\alpha & t \gg \tau_c
\end{array} \right.
\label{eq_scaling}
\end{equation}
where $\langle  R_{\rm in} \rangle $ is the average bubble radius at time $t = 0$ and $\tau_c$ is a time scale characteristic of the coarsening dynamics. 
Experiments have indeed confirmed the 
asymptotic limit \cite{durian_JPHYS, lambert,lambert10,isert}. The scaling 
exponent is $\alpha = 1/2$ in the dry limit ($\phi \rightarrow 1$) and 
$\alpha = 1/3$ in the limit of bubbly fluids ($\phi \rightarrow 0$). There is 
some numerical evidence that $\alpha$ interpolates between these values 
for intermediate $\phi$ \cite{fortuna12}, but to date there have been no 
studies that systematically probe coarsening in the vicinity of the jamming point.

Here, we model coarsening using a variant of the Durian 
bubble model \cite{PhysRevLett.75.4780}. This is a soft sphere model, describing 
the evolution of the spatial configuration of overlapping spheres 
resembling bubbles, droplets or soft particles in a dispersed system. 
Eventhough the model is rather simple, it has been proven very useful in the research of jamming 
in foams and emulsions \cite{liu2010jamming,hecke2010jamming}.
To further advance its capabilities to study coarsening, we have implemented inter-bubble gas 
diffusion as an additional degree of freedom to our set of dynamical 
equations. This is done in the spirit of Gardiner {\textit et al.} \cite{gardiner2000coarsening}, who
studied coarsening in the bubble model at gas fractions far above the jamming limit.
At this range the model  successfully reproduced the asymptotic limit, Eq.~\ref{eq_scaling}, with 
the appropriate scaling of the average radius.

The present work represents the first numerical study of coarsening in the 
bubble model near jamming.  
We  present several main results.  First, the value of the critical volume 
$\phi_c$ is altered by the coarsening dynamics. 
We find that aged samples are particularly efficient at filling voids, and 
$\phi_c$ increases to the unusually large value of $0.87$ in 2D. 
Next, near jamming the scaling state  of Eq.~(\ref{eq_scaling}) holds, 
with an exponent $\alpha \approx 0.45$ and a coarsening time $\tau_c$ 
that diverges on approach to $\phi_c$. 
Finally, coarsening dramatically influences mechanical response near jamming. 
We characterize the storage modulus and the viscous relaxation time, 
both of which scale with sample age and the distance to $\phi_c$.

\section{Bubble model with coarsening}
\begin{figure}[h]
\begin{center}
\includegraphics[width=\columnwidth]{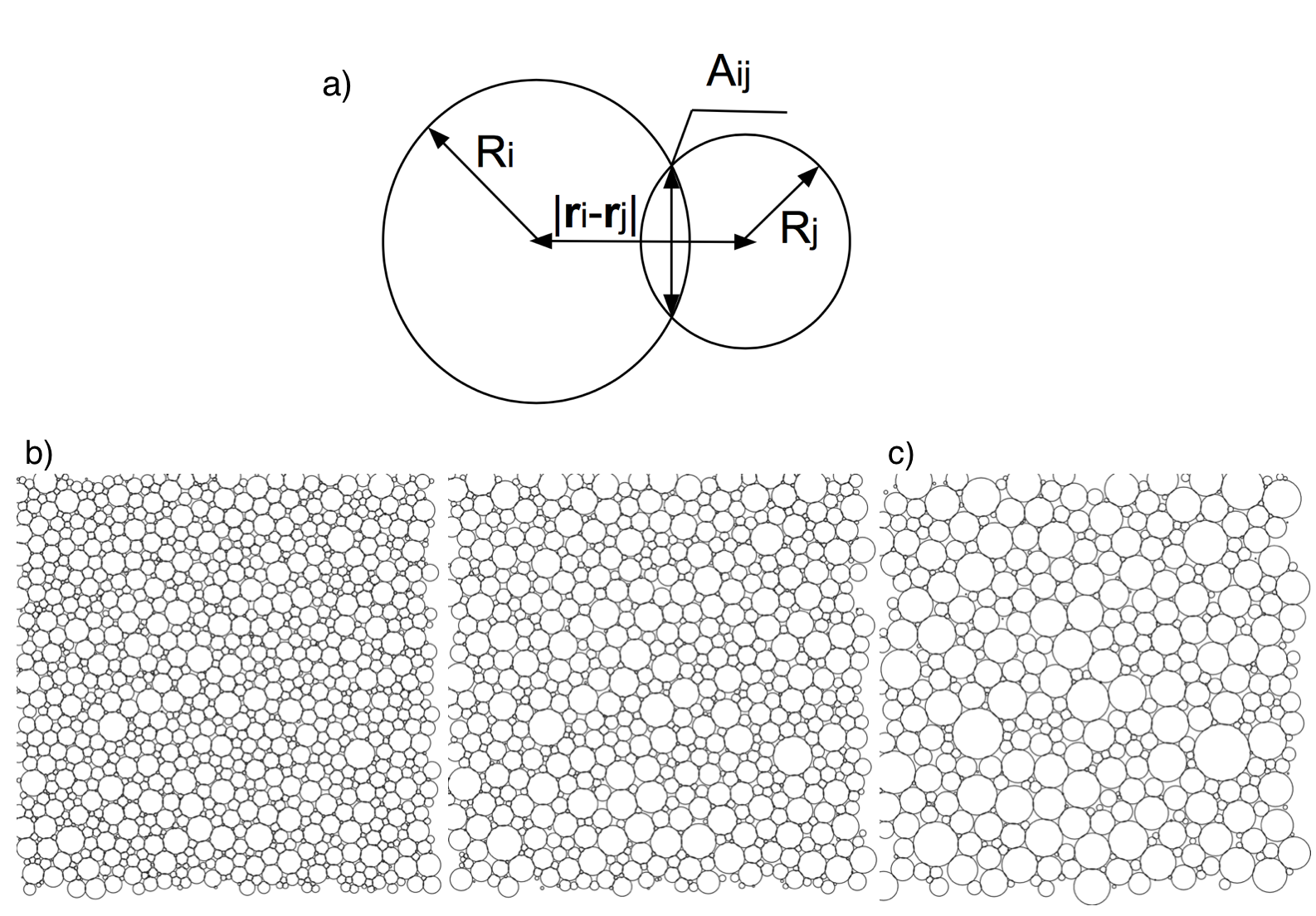}
\caption{\label{fig_bubble_structure}a) A schematic illustration summarizing the essential parameters of the model incorporating the interaction between two overlapping soft spheres. b) and c) are visualizations of a series of snapshots of the simulation: b) shows the initial structure before coarsening, and an intermediate stage, and c) the structure in the scaling state. }
\end{center}
\end{figure}

We model foams using Durian's bubble model \cite{PhysRevLett.75.4780,PhysRevE.55.1739}.
The bubble model describes foams at the bubble level as packings of randomly distributed
soft spheres (or disks in 2D). These spheres interact via a harmonic pair potential proportional to their overlap 
\begin{equation}
\textbf{F}_{ij} = F_0\left(\frac{R_i+R_j-\mid{{\textbf r}_i-{\textbf r}_j}\mid}{R_i + R_j}\right)\frac{{{\textbf r}_i-{\textbf r}_j}}{\mid{{\textbf r}_i-{\textbf r}_j}\mid} \,,
\label{eq:force}
\end{equation}
as illustrated in Fig.~\ref{fig_bubble_structure}a. 
The model assumes the fully overdamped limit: the bubbles are massless, and the force due to bubble overlap
${\textbf F}_i$  must at all times be compensated by the drag force ${\textbf F}^d_i =
-\mu_0 ({\textbf v}_i - \langle  {\textbf v}_j \rangle  )$. Here $\langle  {\textbf v}_j
\rangle $ is the velocity of the background fluid often computed as the average
velocity of the neighboring bubbles, and $\mu_0$  is the
viscosity of the fluid. Since we impose no external deformation
to our system, we set $\langle  {\textbf v}_j \rangle $ = 0.
Then, the bubbles follow a  quasistatic equation of motion with
\begin{equation}
{\textbf v}_i = \frac{1}{\mu_0} \sum_{i \neq j}^N {\textbf F}_{ij},
\label{eq:vel}
\end{equation}
which can be integrated in a molecular dynamics fashion to obtain the evolution of the foam.
From considerations of dimensionality assume that  $F_0 = \sigma_0 \langle R_{\rm in}\rangle $. In this case, the dynamics of the system is determined by ratio $\mu_0/\sigma_0$. 
Therefore, the time scale in the simulation can be chosen so that Eqs.~\ref{eq:force} and~\ref{eq:vel} become
\begin{equation}
{ \frac{d\textbf{x}_i}{dt}=  \sum_{i \neq j}^N { {\langle  R_{\rm in} \rangle }\left(\frac{R_i+R_j-\mid{{\textbf r}_i-{\textbf r}_j}\mid}{R_i + R_j}\right)}  \frac{{{\textbf r}_i-{\textbf r}_j}}{\mid{{\textbf r}_i-{\textbf r}_j}\mid},}
\label{eq:motion}
\end{equation}
where $t$ is dimensionless time, scaled with $\mu_0/\sigma_0$.
The main merits of the model are that it is sufficiently simple, while it still allows to
easily vary the foam properties such as polydispersity, volume fraction, and dimensionality. 

In coarsening, gas diffuses from smaller bubbles to larger ones driven by the
difference of their respective Laplace pressures.
To take this into account in the bubble dynamics model, Gardiner, Dlugogorski, and 
Jameson (henceforth GDJ)  proposed a scheme where, in addition to their elastic and 
viscous interactions, the bubbles are allowed to exchange gas 
\cite{gardiner2000coarsening}. The gas exchange rate of two contacting bubbles is 
proportional to their interaction area $A_{ij}$
and the difference in their Laplace pressures. Each bubble's volume $V_i$ 
changes according to
\begin{equation}
\frac{dV_i}{dt} = K A_{ij} \left( \frac{1}{R_j} - \frac{1}{R_i} \right) \,,
\label{eq_flux}
\end{equation}
where $K$ is the diffusion parameter, encapsulating the properties of the 
liquid film and the
bubbles, such as the effective permeability, surface tension, and 
temperature. 
We integrate this set of differential equations using a second 
order adaptive step size predictor-corrector scheme with error tolerances 
set to $1\cdot10^{-6}$. 

The simulation procedure is as follows.
The simulations begin by first randomly distributing 3000 bubbles in a
periodic rectangle (2D) at the initial volume fraction of $\phi= 0.45$. For
each bubble an initial radius is assigned according to a Gaussian distribution with
the mean $\langle  R_{\rm in} \rangle  = 0.006$ and variance $21\%$. To reach the target volume 
fraction $\phi_0$ we then compress the structure by rescaling the dimensions of the 
simulation cell at a constant velocity.
After the compression, we equilibrate the system by allowing the bubble positions to
relax until the energy changes less than 0.0001\% for 1000 iterations.
This gives us an initial structure, such as the one shown in
 Fig.~\ref{fig_bubble_structure}b. 
Finally, we turn on the coarsening and run the simulations until the number of
bubbles is smaller than a cut-off, which we take to be 300, and the bubble size 
distribution has reached
the scaling state (Fig.~\ref{fig_bubble_structure}c). In all the plots involving time, 
we have used the time scale $t^* = \frac{K}{\langle  R_{\rm in} \rangle ^2}t$.
\begin{figure}
\centering
\begin{tabular}{cc}
\subf{\includegraphics[width=\columnwidth/2]{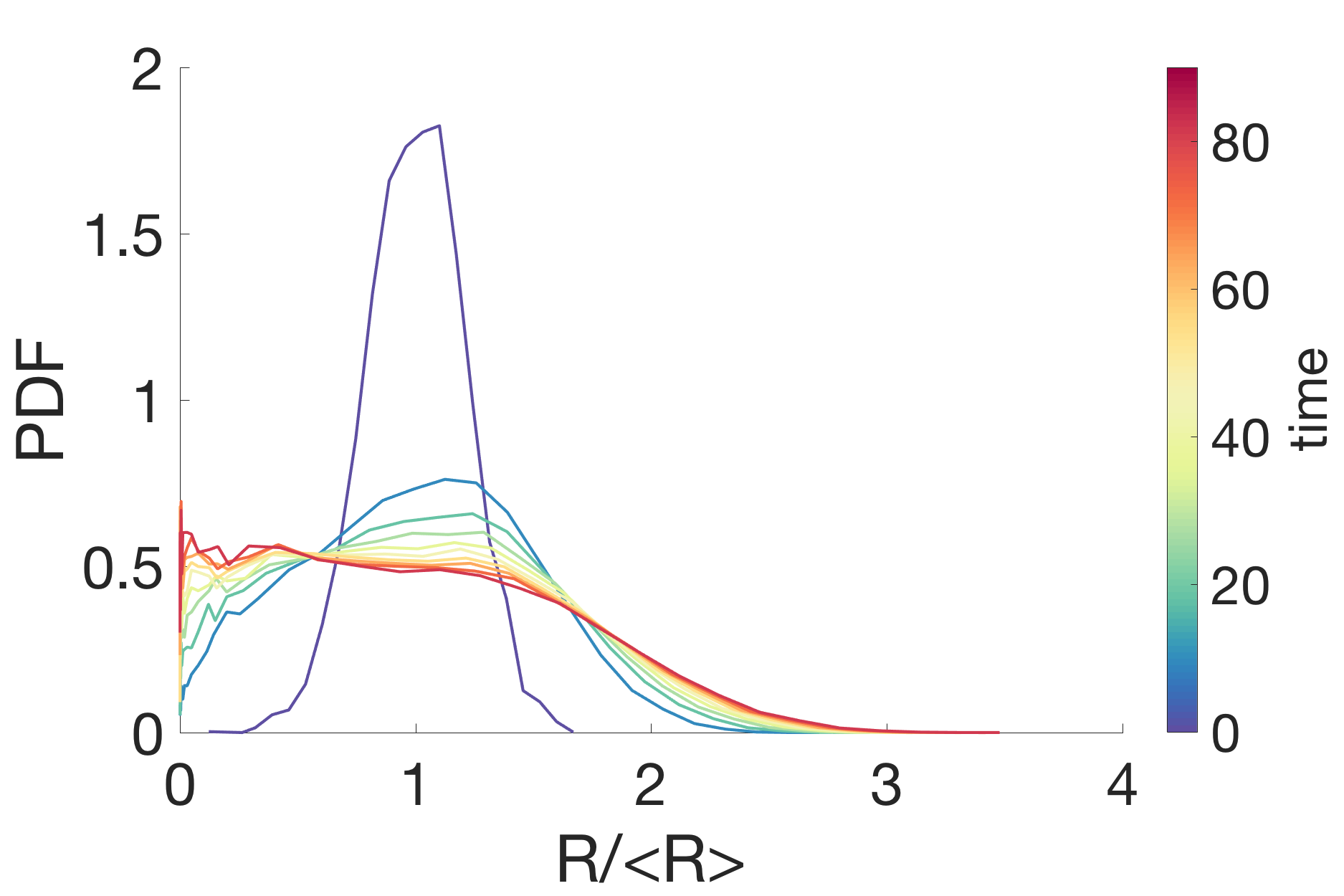}}
     {$\phi = 0.86$}
&
\subf{\includegraphics[width=\columnwidth/2]{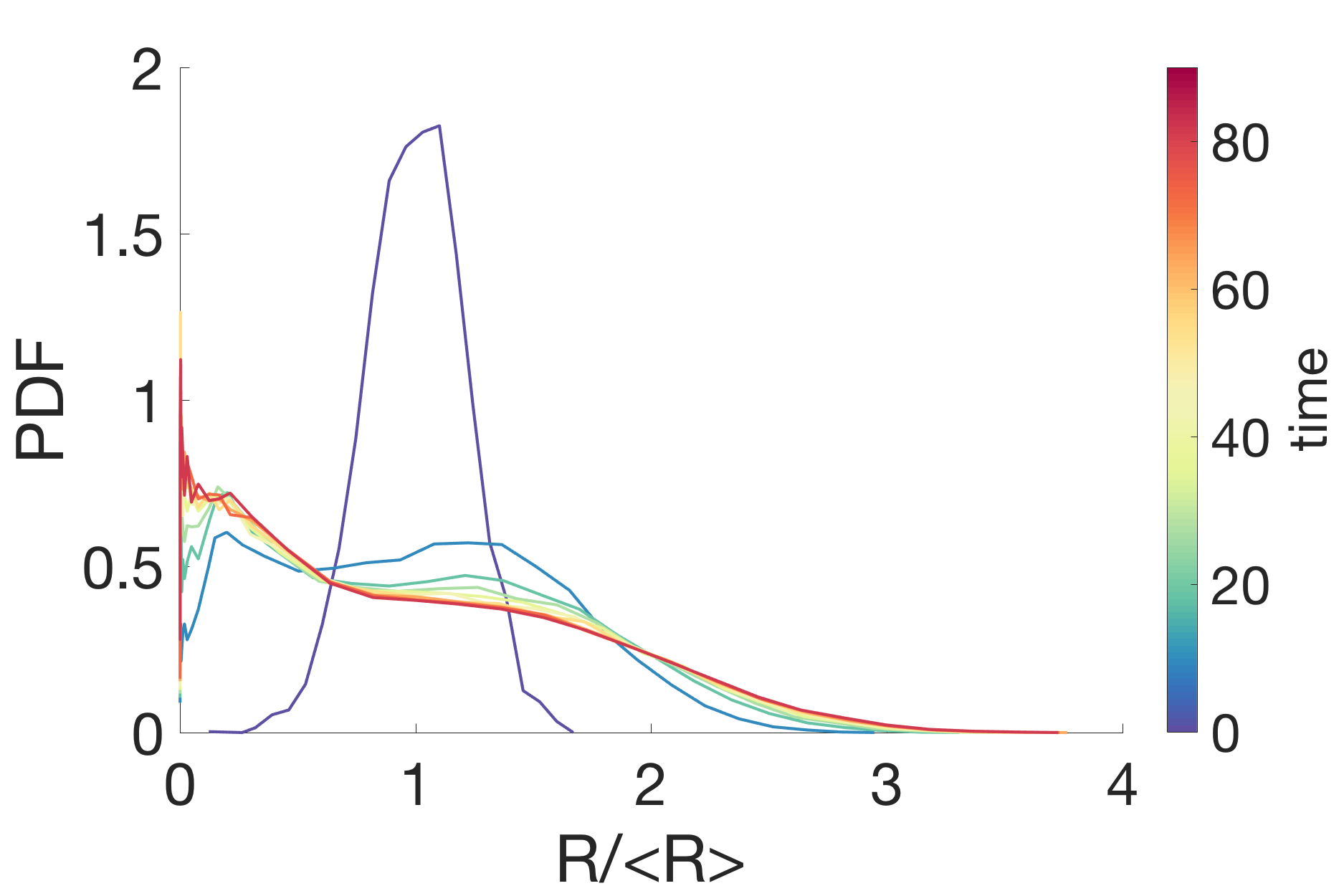}}
    {$\phi = 0.87$}
\\
\subf{\includegraphics[width=\columnwidth/2]{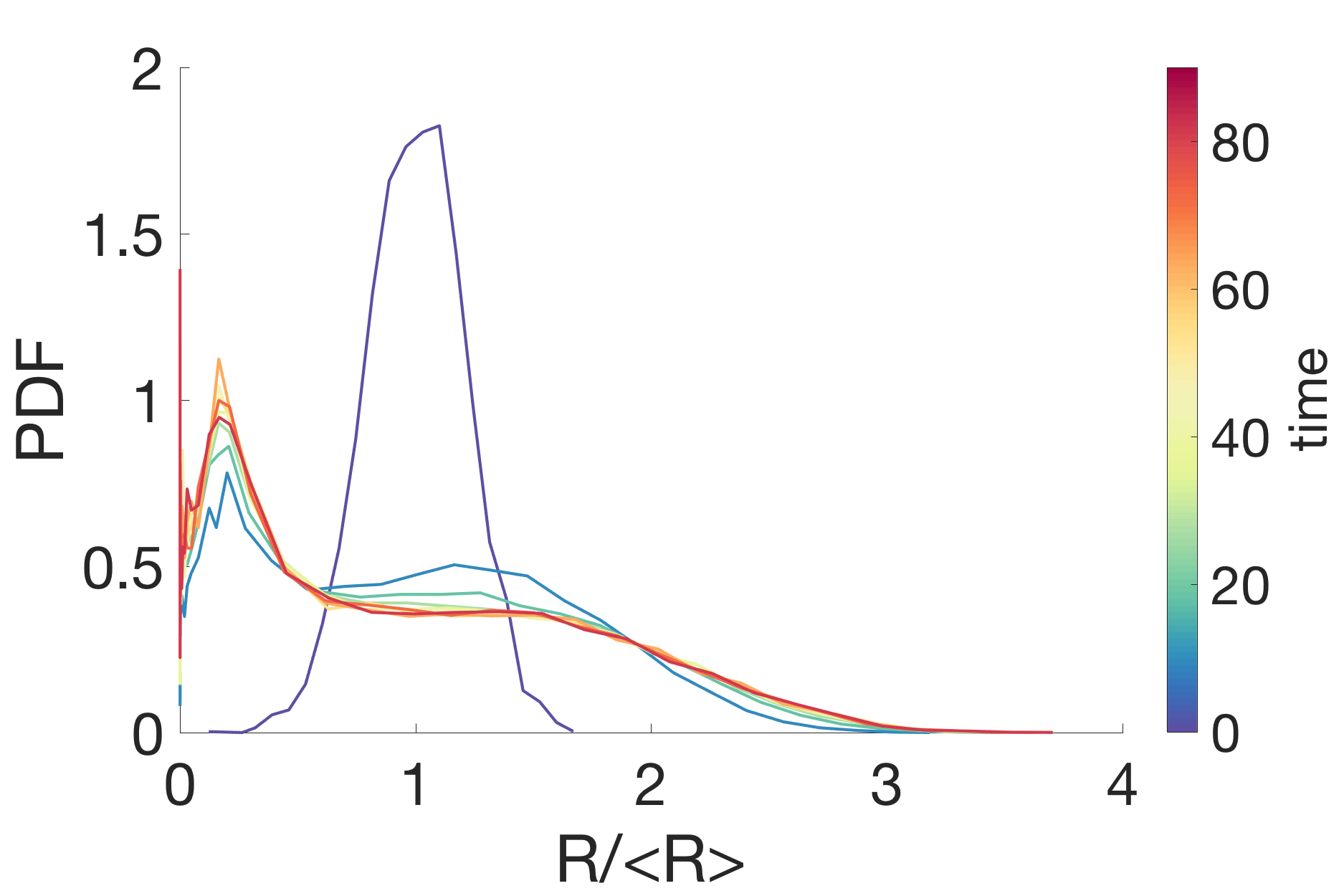}}
   {$\phi = 0.88$}
&
\subf{\includegraphics[width=\columnwidth/2]{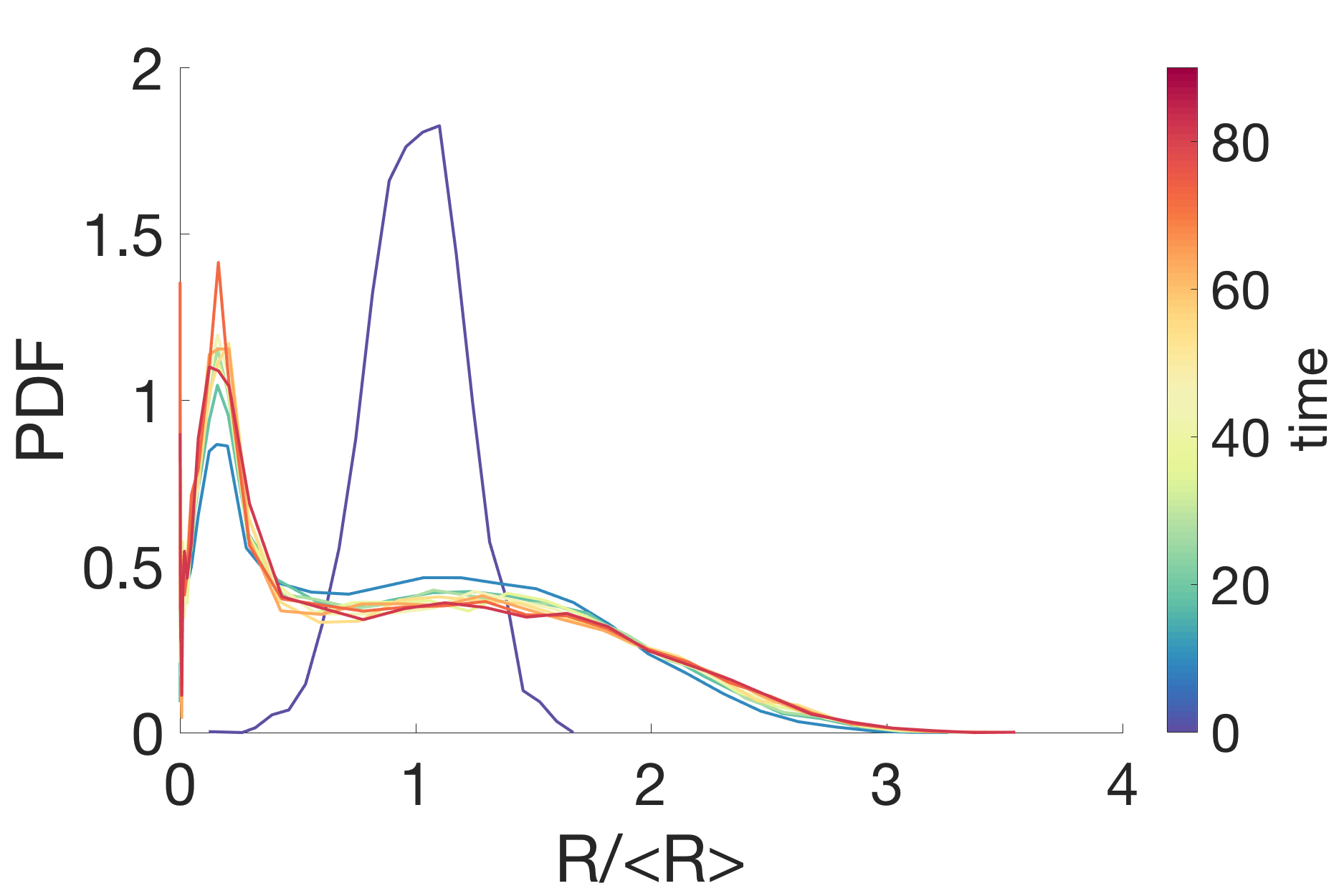}}
    {$\phi = 0.90$}

\end{tabular}
\caption{\label{fig_pdf} The evolution of the probability distribution function (PDF) of the radius (normalized by $\langle  R \rangle $) for different volume fractions ranging from $\phi=0.86$ to $\phi=0.90$. }
\end{figure}

\section{Scaling state}

\begin{figure*}
\centering
\begin{tabular}{cc}
\subf{\includegraphics[width=\textwidth/3]{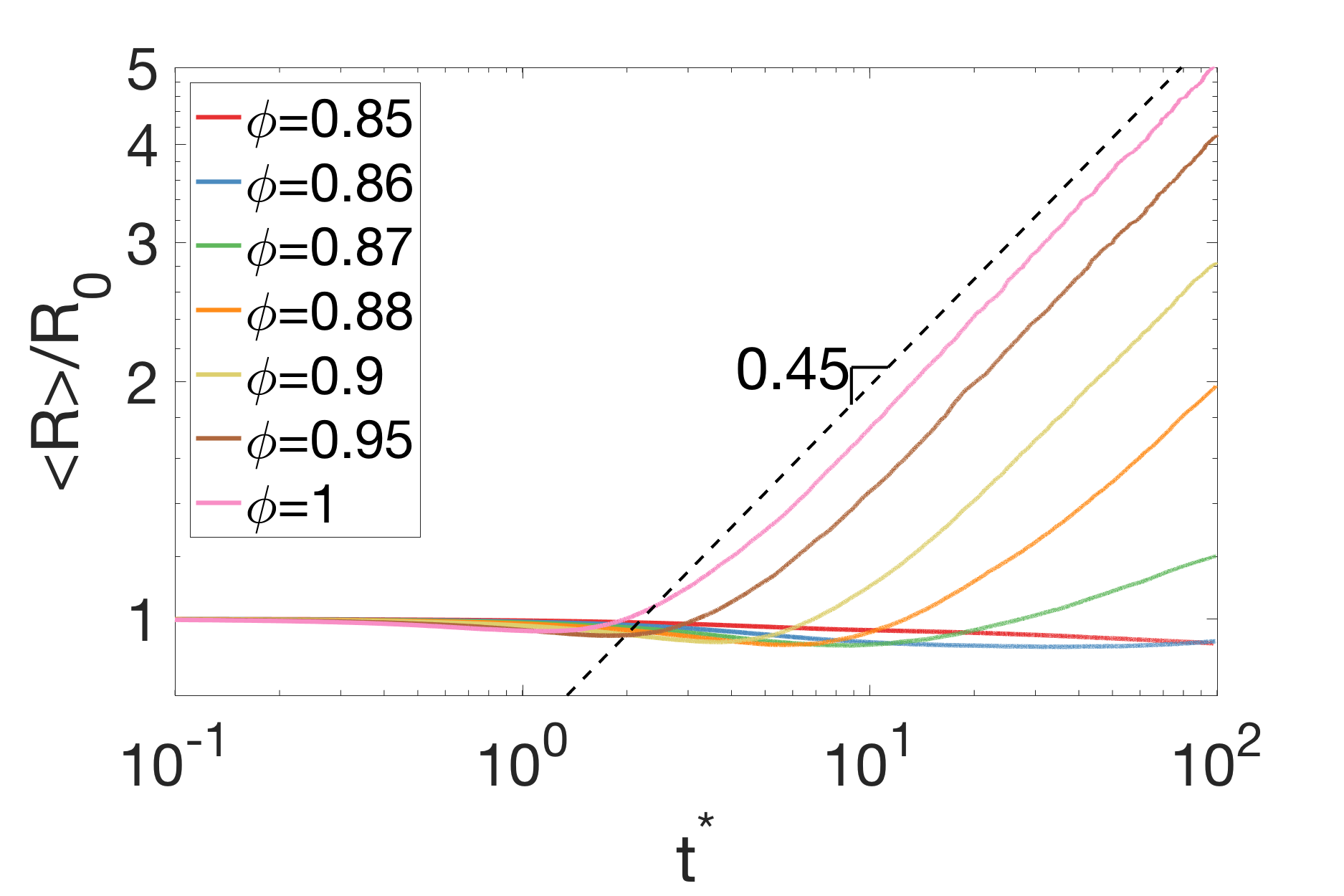}}
     {(a)}
&
\subf{\includegraphics[width=\textwidth/3]{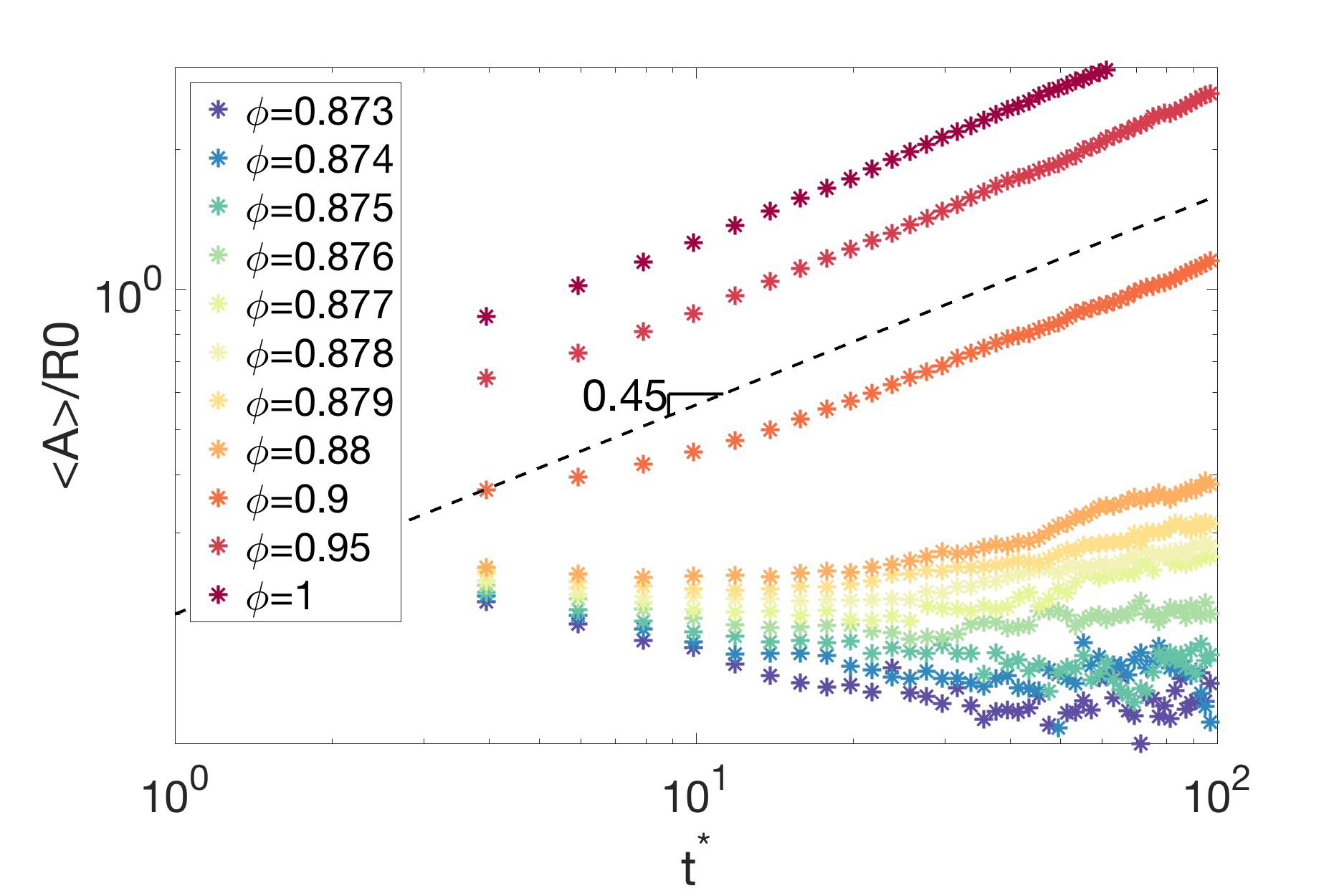}}
	{(b)}

\\
\subf{\includegraphics[width=\textwidth/3]{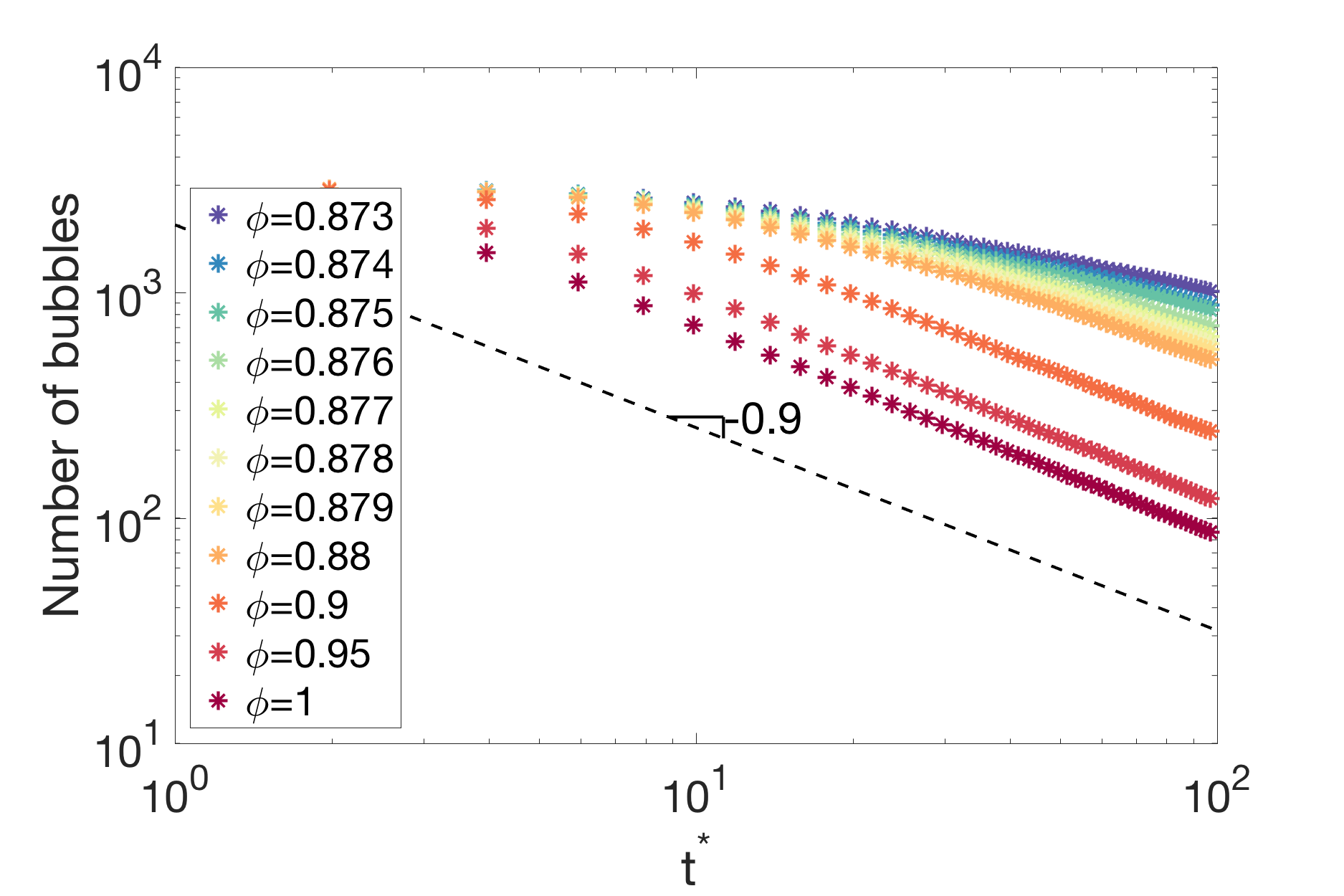}}
	{(c)}
&
\subf{\includegraphics[width=\textwidth/3]{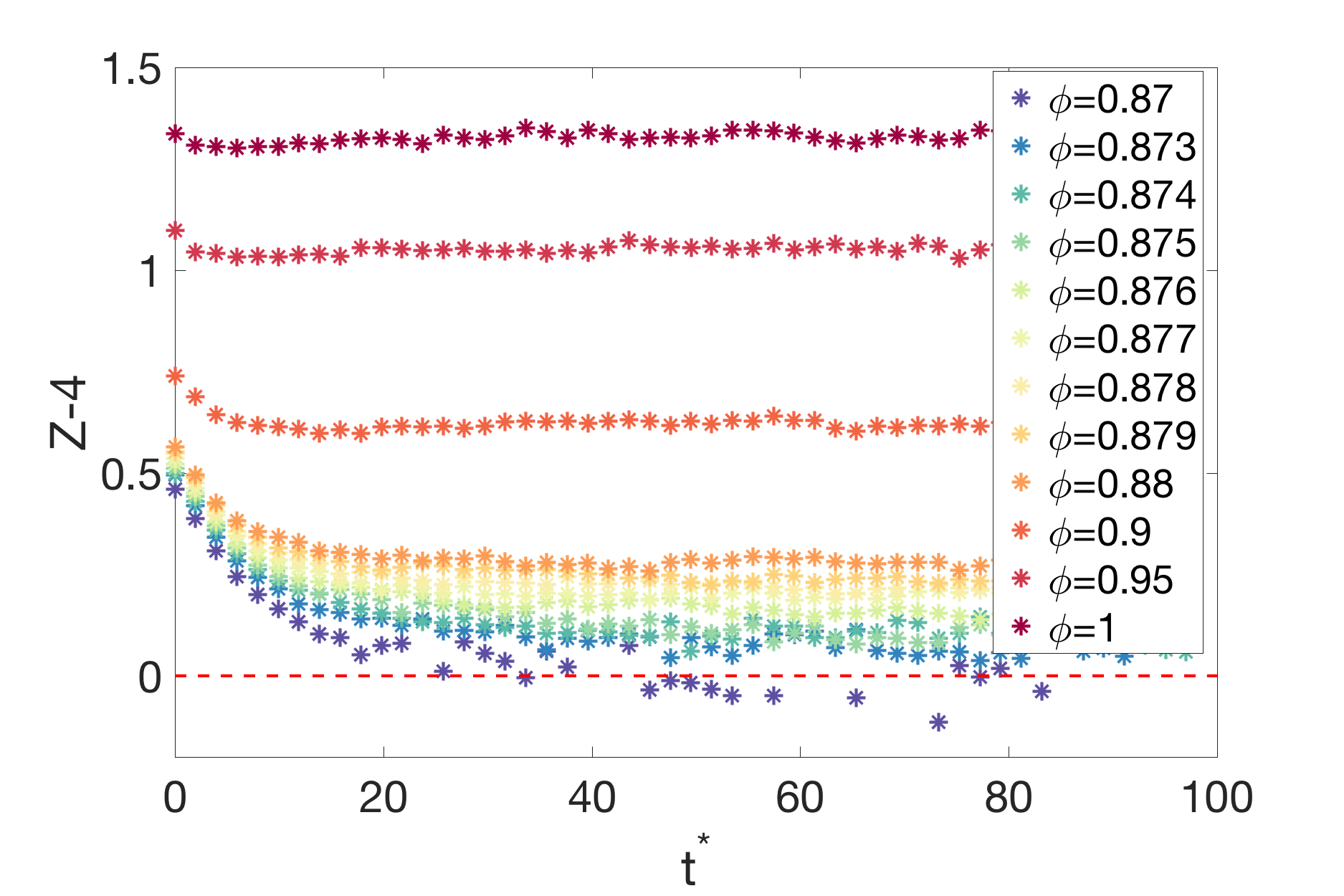}}
    {(d)}
\\
\end{tabular}
\caption{\label{fig_scaling} a) Average radius scaled with its initial value $\langle  R(t^*)\rangle  /R_{\rm in}$ as a function of time $t^*$;  b) Average contact area A scaled with initial average radius as a function of time $t^*$;  c) Total number of bubbles as a function of time $t^*$; d) Mean coordination number $z(t^*)-4$  for varying $\phi$ as a function of time.}
\end{figure*}

During an induction period at the beginning of the coarsening, the probability density 
function (PDF) of  bubble sizes $R/\langle  R(t^*) \rangle $ broadens (Fig.~\ref{fig_pdf}), while 
the average bubble size $\langle  R(t^*) \rangle $ remains approximately constant -- see 
Fig.~\ref{fig_scaling}a. After the induction time the system reaches a scale-independent 
regime where ${\rm PDF}(R/\langle  R(t^*) \rangle )$ ceases to evolve with time (see 
Fig.~\ref{fig_pdf}). 
At each volume fraction, the PDF in the scaling 
state exhibits a significantly broader shape, along with a high number of 
small bubbles of size around $R \approx 0.1 R_{max}$. This most likely 
relates to the way the system optimizes the occupied space by filling 
the voids between the big bubbles with fitting small ones. The small 
bubbles have no overlap with any of their neighbors (are so-called rattlers). 
As a result, having no neighbors,  they will not experience
any gas exchange until their surrounding bubbles change their 
configuration. Furthermore, the rattler radius geometrically allowed in a 
foam increases with the average bubble size.

The average bubble radius in the scaling state follows a power law
 $\langle  R \rangle  \sim {t^*}^{\alpha}$, as observed in Fig.~\ref{fig_scaling}a for
  a range of volume fractions. While not all volume fractions reach the
   scaling state, among those that do we find an exponent 
$\alpha \approx 0.45$. This is strikingly close to the value of $1/2$ required 
by von Neumann's law in perfectly dry foams ($\phi = 1$) -- despite the fact 
that von Neumann's law does not hold exactly in the bubble model. Due to 
dimensional considerations and the conservation of total volume, one also 
anticipates related scaling relations for the mean contact ``area'' 
(length in $D = 2$ dimensions) between bubbles, $\langle  A \rangle  \sim 
t^{\alpha}$, as well as the total number of bubbles, $N \sim {t^*}^{-D\alpha}$. 
These are verified in Fig.~\ref{fig_scaling}b and c.

It is apparent from Fig.~\ref{fig_scaling}a that the onset of the scaling state 
is determined by a volume fraction-dependent time scale $\tau_c(\phi)$. 
The volume fraction $\phi \approx 0.87$ appears to be a marginal case; 
lower values of $\phi$ show no evidence of scaling within the simulation 
runtime $T = 100$, while larger values show clear power law scaling at long 
times. In order to test for connections to the jamming transition, in 
Fig.~\ref{fig_scaling}d we plot the mean coordination number for varying 
$\phi$ as a function of time. In determining $z$, we assume that the 
system's evolution is quasistatic, so that ``rattlers'' can be meaningfully identified and 
removed. Recall that jamming occurs at the critical coordination number 
$z_c = 2D = 4$ in two dimensions, in accord with a constraint counting 
argument that dates to Maxwell. 
One clearly sees that packings for $\phi \ge 0.88$ satisfy $z >  z_c$ for the 
entire runtime, while $\phi = 0.87$ dips below 4 approximately half way 
through the run -- the critical value $\phi_c$ has increased due to 
coarsening. As all initial conditions are jammed, the 
increase in the jamming volume fraction, and the corresponding drop in the 
average coordination number, relate to the change in the bubble size 
distribution under coarsening. When the bubble size distribution reaches 
the scaling state the bubbles fill 
space more optimally, by reducing the bubble overlap via the gas diffusion. 
In order to estimate $\phi_c$ more accurately, we have probed the interval 
$0.87 \le \phi \le 0.88$ in steps of $0.001$ for longer runs to $T = 1000$. At these long times the 
system reaches sizes $N \sim {\cal O}(100)$ (recall that $N \sim 1/{t^*}^{D\alpha}$) and 
the coordination number experiences large fluctuations, which restricts our 
ability to determine a precise 
jamming volume fraction $\phi_c$. We find that for volume fractions $\phi \le 0.872$ the 
mean coordination unambiguously drops below $z_c$ within the runtime, while for $\phi 
\ge 0.877$ it clearly remains above $z_c$. For $0.873 \le \phi \le 0.876$, the coordination 
number fluctuates on either side of $z_c$, suggesting that the system's evolution passes 
through both jammed and unjammed configurations.

\begin{figure}[h]
\begin{center}
\includegraphics[width=\columnwidth]{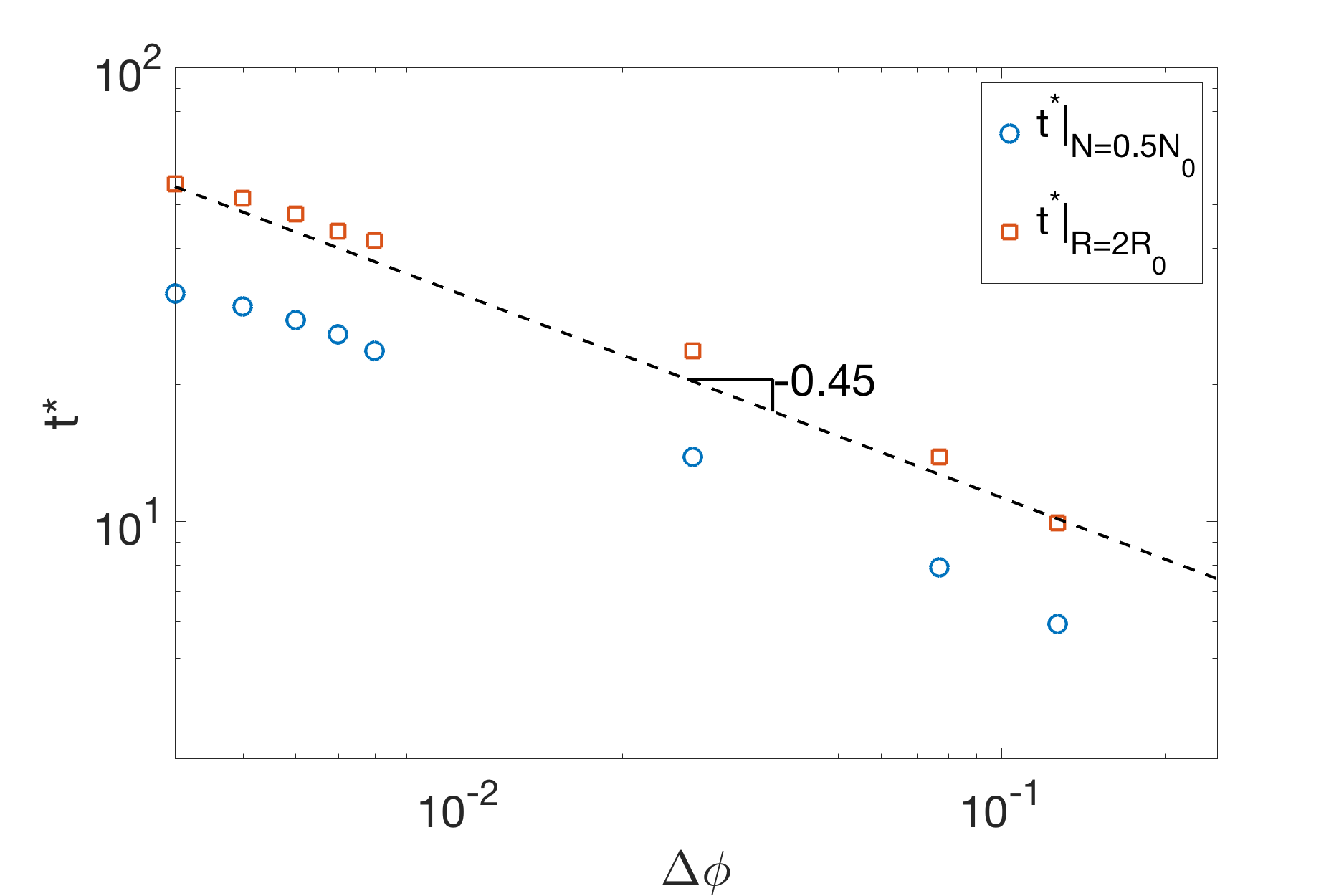}
\caption{\label{fig_Prefactor_phi} Estimation of $\tau_c$. Blue circles indicate the time when the initial number of bubbles decreases by a factor of 2.Red squares indicate time when $\langle  R\rangle /R_{\rm in}$ is equal to 2.}
\end{center}
\end{figure}

The time scale $\tau_c(\phi)$ where the scaling state is reached can be estimated from the 
evolution of the average bubble radius. In Fig.~\ref{fig_Prefactor_phi} we estimate 
$\tau_c$ as the time where $\langle  R\rangle /R_{\rm in}$ is equal to 2, i.e.~the average bubble has increased 
its radius by $100\%$. The same can be done by computing time where initial number of 
bubbles decreased by a factor of 2. The data in both cases are reasonably fit by a power 
law $\tau_c \propto 1/(\phi - \phi_c)^\alpha$, with fitting parameters $\phi_c = 0.873$ and 
$\alpha \approx 0.45$. Deviations at small values of $\Delta \phi \equiv \phi - \phi_c$ are clearly present; we 
expect they are associated with the fluctuations between jammed and unjammed states 
identified above.

\section{Mechanics}

In the previous Section we identified signatures of the jamming transition in the scaling 
dynamics of coarsening foams. We now seek to correlate these effects with features of 
the mechanical response.

Bubbles store energy when distorted and dissipate energy when sliding past each other, 
which gives rise to viscoelastic response.\cite{durian95,hohler05} 
Foam viscoelasticity can be probed both experimentally and numerically by measuring the 
complex shear modulus $G^*(\omega) = G'(\omega) + \imath G''(\omega)$, defined as the 
complex ratio of shear stress and strain amplitude under harmonic forcing at angular 
frequency $\omega$.\cite{barnes} The real and imaginary parts of $G^*$ are known as the storage and 
loss moduli, respectively. For reference, we recall that simple viscoelastic solids 
(frequently referred to as Kelvin-Voigt solids) have a constant storage modulus $G_0$ 
and a linear loss modulus $\eta_0 \omega$; their ratio $G_0/\eta_0$ selects a 
characteristic relaxation time. Deviations from the Kelvin-Voigt form are associated with a 
spectrum of relaxation times (rather than a single time scale), but the time scale $\tau_r$ 
where the loss and storage moduli cross, i.e.~$G'(1/\tau_r) = G''(1/\tau_r)$ is still an 
important reference point. It indicates a crossover from predominantly solid-like response 
at low frequencies to predominantly liquid-like response at high frequencies.

While the complex shear modulus is meant to describe steady state dynamics, the 
structure of coarsening foams in the scaling state continuously evolves (``ages'').\cite{cohenaddad98,gopal03} 
Aging violates time translational invariance, which is assumed in the usual definition of the complex shear modulus.\cite{fielding00} Nevertheless, oscillatory rheology is often used to characterize soft materials, with a focus on frequencies that are fast compared to the evolution of the structure.\cite{cohenaddad98,gopal03}
Here we determine the storage and loss moduli via a numerical ``experiment'' that 
disentangles viscous relaxation from coarsening-induced rearrangements. States are 
sampled at varying times $t^*$ from a coarsening simulation at fixed $\phi$. Coarsening 
dynamics are then turned off (all particle sizes are held fixed) and the complex shear 
modulus is measured according to the method described below. In this way it is possible 
to obtain moduli over the full range of frequencies $0 \le \omega \le \infty$; however, we 
focus on time scales that are short compared to the system's age.

In order to measure the complex shear modulus, we employ a linearization scheme introduced in Ref.~\cite{tighe11}. 

Given a particular configuration of bubbles, its collective response to shear is described 
by the $DN+1$-component vector ${\textbf u} = (\vec u_1, \vec u_2, \ldots \vec u_D, 
\gamma)^T$, where $\vec u_i$ is the displacement of bubble $i$ from its reference 
position, and $\gamma$ is the shear strain experienced by the unit cell. The response to a 
shear stress $\sigma$ is given by the solution to the first order differential equation

\begin{equation}
{\textbf K} \, {\textbf u}(t^*) + {\textbf B} \, \dot {\textbf u}(t^*)  = \sigma(t^*) V {\rm \textbf e}_\gamma\,,
\label{eqn_viscoelasticity}
\end{equation}
where ${\rm \textbf e}_\gamma$ is a unit vector along the strain coordinate. The stiffness 
matrix ${\textbf K}$ and damping matrix ${\textbf B}$ describe the elastic and viscous forces on 
the particles, respectively. ${\textbf K}$ consists of second derivatives of the elastic potential 
energy with respect to the particle and strain degrees of freedom; ${\textbf B}$ is similarly 
defined in terms of the Rayleigh dissipation function. Details are available in Ref.~\cite{tighe11}. 
Linearization is strictly valid only when deformation amplitudes are 
infinitesimal; nevertheless, numerical studies indicate that moduli calculated in this way 
remain accurate over a finite strain interval \cite{vandeen14,vandeen16,boschan16}. 
By Fourier-transforming Eq.~\ref{eqn_viscoelasticity} and solving for the 
complex shear strain in response to a sinusoidally oscillating shear stress
 with frequency $\omega$, one 
can determine complex shear modulus.

\begin{figure}[tb]
\centering

\subf{\includegraphics[width=\columnwidth]{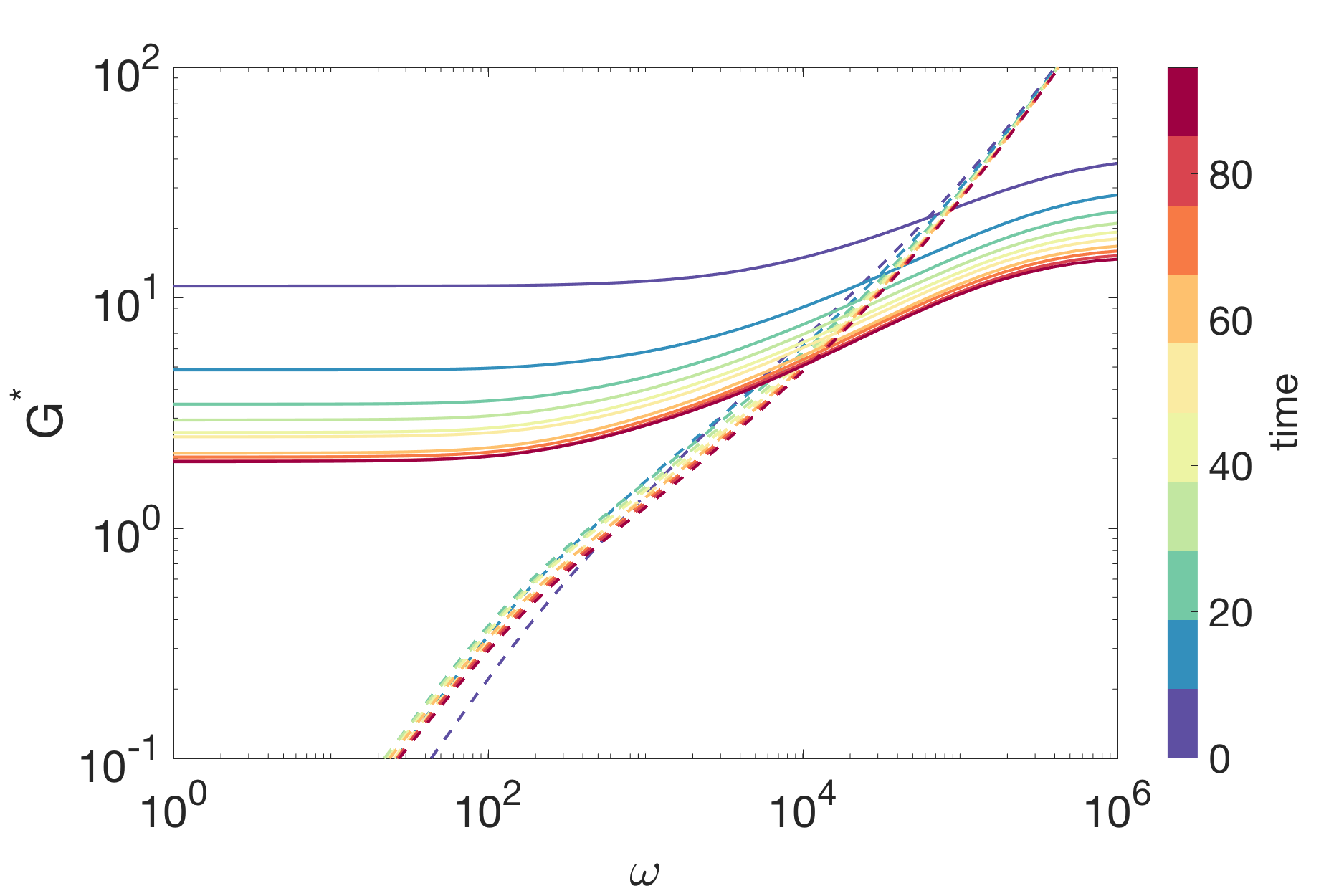}}

\caption{\label{fig_G_star}{Complex shear modulus $G^*$  for $\phi = 0.878$. Solid lines depict the storage modulus $G'$; dashed lines depict the loss modulus $G''$. The color palette corresponds to evolution in time.}}

\end{figure}

In Fig.~\ref{fig_G_star}, we plot the storage and loss moduli (solid and dashed curves, respectively) for a system at $\phi = 0.878$, close to but above the jamming transition. The storage modulus displays a low frequency plateau, indicating that the sampled configurations are jammed solids. (We stress again that the linearization scheme employed here ``turns off'' coarsening, hence any softening at asymptotically low frequencies due to coarsening-induced rearrangements will not be captured.) At high frequencies there is a second plateau in $G'$, associated with affine deformations \cite{tighe11,boschan16}. There is a gradual crossover between these two plateaus. In previous work it was shown that this crossover occupies a widening window in frequency as $\phi \rightarrow \phi_c^+$, such that $G' \sim G'' \sim \omega^{1/2}$.\cite{tighe11}  The loss modulus $G''$ is comparatively simple; it is nearly linear over the entire frequency range, consistent with a Kelvin-Voigt solid. Experimental measurements of the loss modulus in foams often show a plateau at low frequency.\cite{cohenaddad98,gopal03} This feature is absent from our data, and from prior studies of Durian's bubble model without coarsening.\cite{tighe11} We suggest that the plateau results from physics that is not incorporated in the bubble model, e.g.~due to thin film flow.

\begin{figure*}[htb]
	\centering
	\begin{tabular}{cc}
		\subf{\includegraphics[width=\textwidth/2]{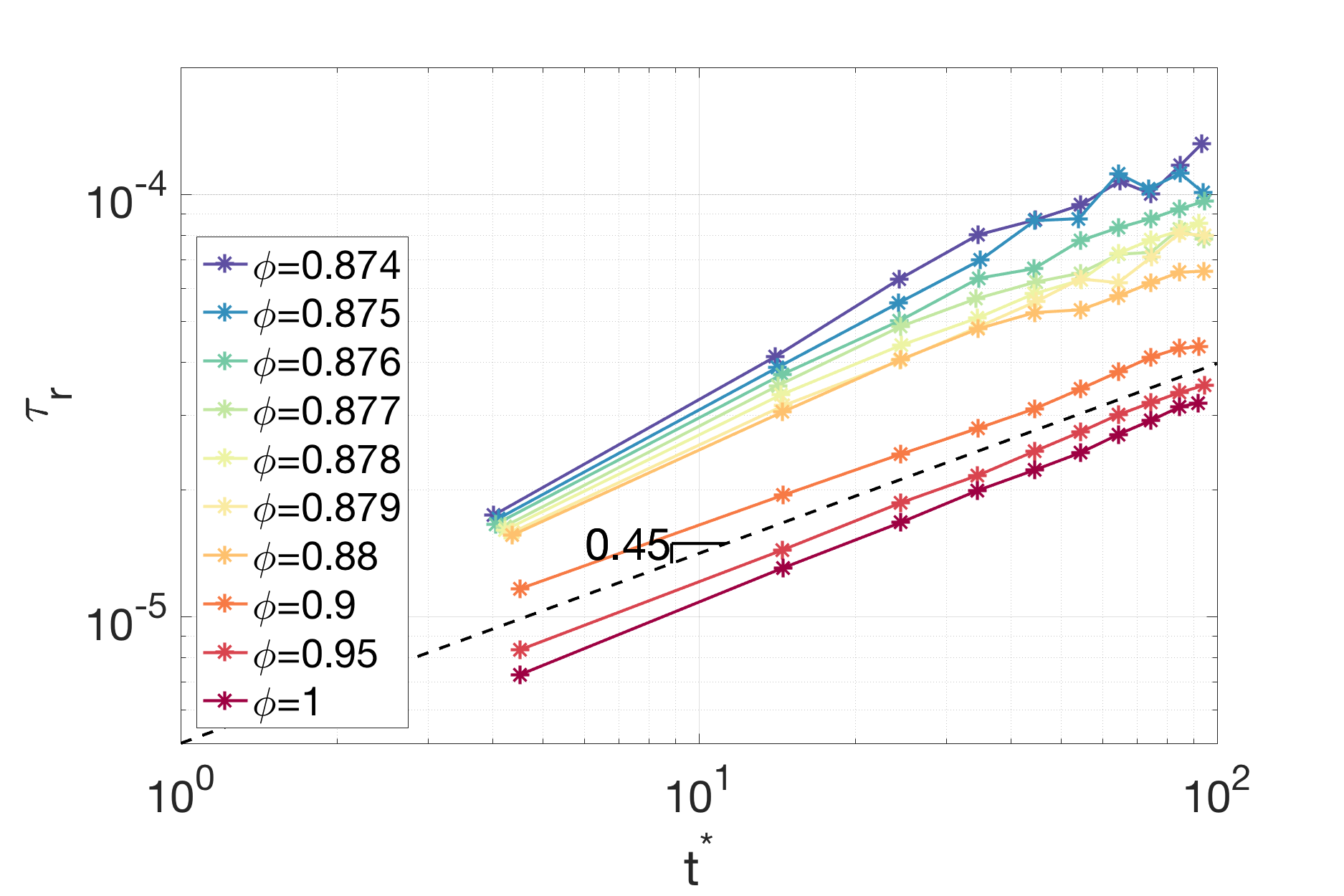}}
		{(a)}
		&
		\subf{\includegraphics[width=\textwidth/2]{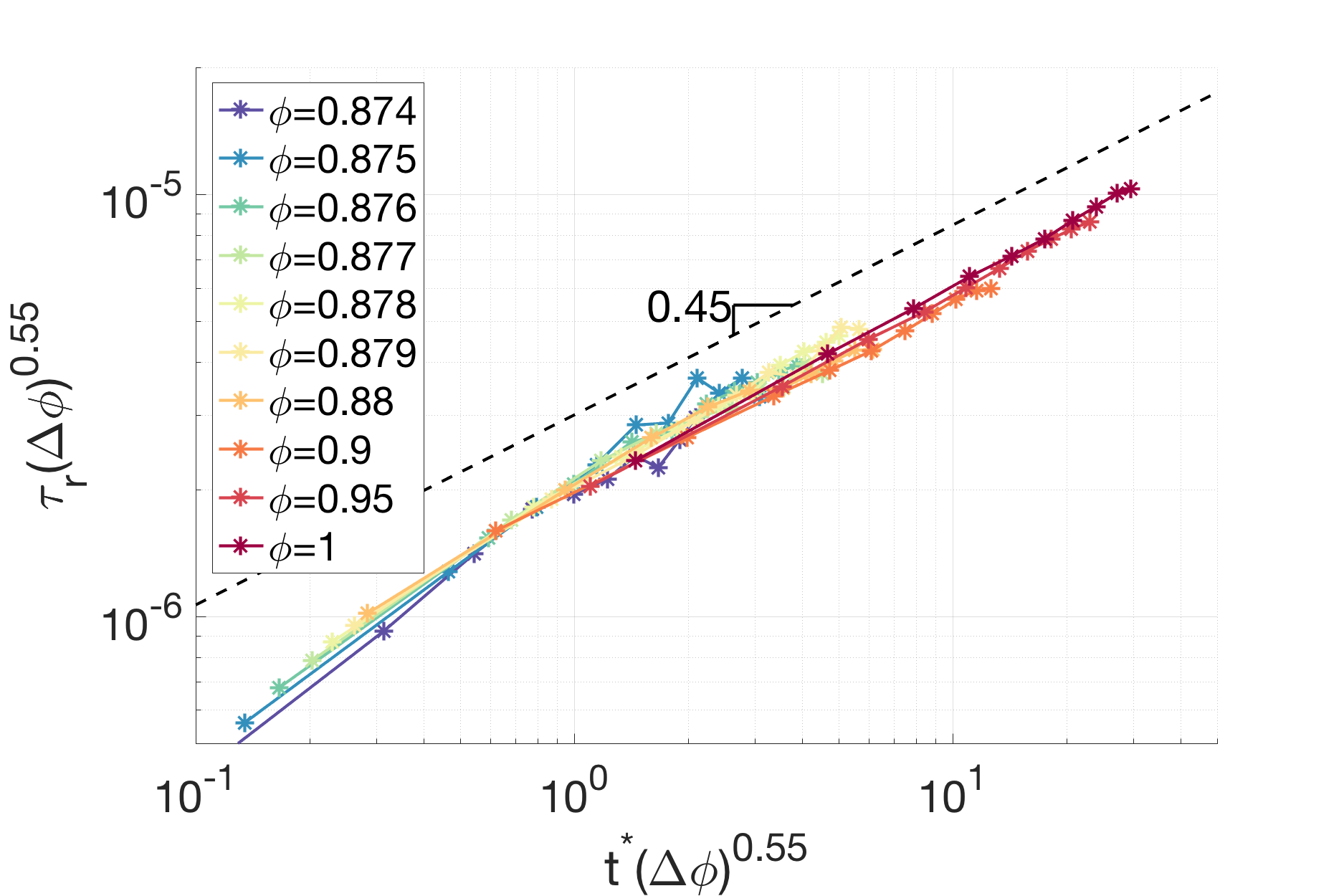}}
		{(b)}
		\\
	\end{tabular}
	\caption{\label{fig_tau} (a) Mechanical relaxation time measured from the intersection of $G'$ and $G''$ for different $\phi$ as a function of system age. (b) Data collapse: relaxation times scaled with $\Delta\phi^{0.55}$ as a function of $t^*\Delta\phi^{0.55}$.}
\end{figure*}

The storage modulus shows a clear dependence on the sample age, with an overall downward shift with increasing age. Their intersection point defines
the mechanical relaxation time $\tau_r$, which we measure at varying volume fraction and age -- see Fig.~\ref{fig_tau}a. 
The relaxation time clearly depends on both 
the system age $t^*$ and the distance to jamming; it grows larger with increasing age and decreasing  distance to jamming $\Delta \phi$. The dependence of $\tau_r$ on $t^*$ and $\Delta \phi$ can be rationalized in two steps, beginning with its growth with age.

The age dependence of $\tau_r$ is controlled by the scaling of $G'$ and $G''$ with $t^*$. These can be anticipated on dimensional grounds, by which one expects there to be characteristic elastic and viscous stress scales 
\begin{align}
\sigma_{\rm el} &\sim \left(\frac{N(t^*) \, R(t^*)}{L^D}\right) F_{\rm el}(t^*) \\
\sigma_{\rm visc} &\sim \left(\frac{N(t^*) \, R(t^*)}{L^D}\right) F_{\rm visc}(t^*) \,.
\end{align}
Here $N$, $R$, $F_{\rm el}$, and $F_{\rm visc}$ are typical values of the 
particle number, particle radii, and elastic and viscous forces, respectively; 
$L^D$ is the volume of the unit cell, which is constant. The elastic force 
law scales with the dimensionless overlap (c.f.~Eq.~\ref{eq:force}) and 
should therefore be independent of time in the scaling state; hence the 
time-dependence of the typical elastic stress scales as $\sigma_{\rm el} 
\sim N(t^*) R(t^*) \sim {t^*}^{-\alpha}$. By contrast, the typical viscous force is set 
by the bubble velocity $V \propto R(t^*) \, \omega$. 
Hence $\sigma_{\rm visc} \sim {t^*}^0$, consistent with observations. 
Turning back to $\tau_r$, we note that the relaxation time in systems without coarsening 
is insensitive to the distance to jamming; hence the dependence here is 
likely to be inherited from the coarsening dynamics. And indeed, a simple 
balancing of the viscous and elastic stress scales would suggest a 
frequency $\omega_r \equiv 1/\tau_r$ that depends on age as $\omega_r 
\sim R \sim {t^*}^\alpha$, consistent with the data in Fig.~\ref{fig_tau}a.

In order to understand the dependence of $\tau_r$ on $\Delta \phi$, we postulate that the coarsening time $\tau_c$ sets the natural units for both relaxation time and the age of the system. This is expressed most naturally in the form of a scaling ansatz, 
\begin{equation}
\frac{\tau_r}{\tau_c} \sim 
{\cal T}\left(
\frac{t^*}{\tau_c}
\right) \,,
\label{eqn_rescale}
\end{equation}
for some function ${\cal T}(x)$. Indeed, in Fig.~\ref{fig_tau}b we obtain good data collapse when plotting $\tau_r \, \Delta \phi^{\alpha}$ versus $t^* \, \Delta \phi^\alpha$. Treating $\alpha$ as a free parameter, the best collapse is found for $\alpha = 0.55$, close to the value $0.45$ determined independently above. (Good data collapse for $\phi <  0.95$ can also be obtained using $0.45$.)
As expected, ${\cal T} \sim x^\alpha$ for large values of $x$, when 
the system's age is large compared to the coarsening time $\tau_c$. It 
follows that, in the scaling state, the mechanical relaxation time obeys 
$\tau_r \sim {t^*}^\alpha/\Delta \phi^{\alpha(1 -  \alpha) }$. Deviations from 
a slope of $\alpha$ occur when the age is smaller than the coarsening 
time.

\section{Conclusions}
\balance
We have presented the results of numerical simulations of the coarsening of foams close to, and slightly above, the jamming volume fraction.
For this purpose, we implemented the Durian bubble model, with extensions to incorporate gas diffusion between the bubbles.

Our main observation are: (i) The model captures the expected ${t^*}^\alpha$ scaling, with $\alpha \approx 0.5$, of the average radius well above the jamming limit.
(ii) The critical volume fraction where jamming occurs is shifted to 
$\phi_c \approx 0.873$, significantly higher than that found, e.g., in
typical bidisperse mixtures \cite{koeze16}.
(iii) Approaching $\phi_c$, the characteristic coarsening time $\tau_c$
diverges as $1/\Delta \phi^{\alpha}$.
(iv) The foam's mechanical response is radically influenced by the 
coarsening; the mechanical relaxation time $\tau_r$, where the oscillatory 
rheology shows a crossover from liquid- to solid-like behavior, shows 
scaling both with the system's age and $\Delta \phi$.
Points (ii-iv) are all related to the change of the bubble distribution due to 
coarsening, which increases the jamming volume fraction $\phi_c$, 
unjamming the system partially or completely at times. The final point 
establishes a clear connection between mechanical relaxation and the 
coarsening dynamics.

Finally, our results suggest directions for future work. 
In order to more deeply understand the enhanced packing efficiency of the scaling state, it would be necessary to model the form of the bubble size distribution directly. 
There may be fruitful connections to systems undergoing rupture 
and/or Apollonian packings.\cite{einavPRL2010,anishchikPRL1995} 
An additional question concerns the dominant mechanism of gas exchange between bubbles. 
In the present simulations, fluxes are between pairs of 
particles in contact. 
Obviously this slows down as the gas fraction decreases. 
In bubbly liquids, by contrast, gas exchange is mediated by the fluid. 
A third possibility is that, close to the jamming volume fraction, 
there is a non-negligible flux through the Plateau borders (i.e.~through 
the packing's fluid-filled voids, rather than through the increasingly 
smaller thin film interfaces \cite{durianAPS,cohen2013flow}. 
This would yield additional terms in Eq.~\ref{eq_flux}.
Additionally, in technological applications foams often undergo shear flow during 
coarsening; how are these two forms of driving coupled? 
In addition, industrial foams are often formed of thixotropic complex fluids, such as (nano)particulate suspensions. 
Coarsening dynamics changes due to the non-linear dynamics of the 
suspending liquid, ultimately stopping completely\cite{Britan200915,cohen2013flow} -- how can coarsening and mechanics be characterized and modeled in such cases? 
Many of these issues and questions can potentially be addressed with 
straightforward extensions of the present model.

\section{Acknowledgements}
A.P.~and K.K.~are grateful to the Academy of Finland for financial support through project No. 278367. K.B.~and B.P.T.~acknowledge funding from the Netherlands Organization for Scientific Research (NWO).

\bibliography{master} %
\bibliographystyle{rsc}
\clearpage

\end{document}